\documentstyle[psfig,pre,multicol,aps]{revtex}

\begin{document}

\draft

%%%%%%%%%%%%%%%%%%%%%%%%%%%%%%%%%%%%%%%%%%%%%%%%%%
\title{Curvature-induced symmetry breaking \\
in nonlinear Schr{\"o}dinger models}
%%%%%%%%%%%%%%%%%%%%%%%%%%%%%%%%%%%%%%%%%%%%%%%%%%

%%%%%%%%%%%%%%%%%%%%%%%%%%%%%%%%%%%%%%%%%%%%%%%%%%
\author{Yu.B. Gaididei, S.F. Mingaleev} 
\address{Bogolyubov Institute for Theoretical Physics, 
03143 Kiev, Ukraine}
%%%%%%%%%%%%%%%%%%%%%%%%%%%%%%%%%%%%%%%%%%%%%%%%%%
\author{P.L. Christiansen}
\address{Department of Mathematical Modelling, 
The Technical University of Denmark, 
DK-2800 Lyngby, Denmark}
%%%%%%%%%%%%%%%%%%%%%%%%%%%%%%%%%%%%%%%%%%%%%%%%%%

\date{\today}
\maketitle

%%%%%%%%%%%%%%%%%%%%%%%%%%%%%%%%%%%%%%%%%%%%%%%%%%
\begin{abstract}
We consider a curved chain of nonlinear oscillators and 
show that the interplay of curvature and nonlinearity leads to 
a symmetry breaking when an asymmetric stationary state becomes 
energetically more favorable than a symmetric stationary state. 
We show that the energy of localized states decreases with increasing 
curvature, i.e. bending is a trap for nonlinear excitations. 
A violation of the Vakhitov-Kolokolov stability criterion 
is found in the case where the instability is due to the softening 
of the Peierls internal mode. 
\end{abstract}
%%%%%%%%%%%%%%%%%%%%%%%%%%%%%%%%%%%%%%%%%%%%%%%%%%
%\pacs{}

\begin{multicols}{2}
\narrowtext

Nanoscale physical systems such as biological macromolecules, nanotubes, 
microtubules, vesicles, electronic and photonic wave-guide structures 
\cite{dna,nan,mict,amp,imry,souk} are usually objects with nontrivial 
geometry. 
The gene transcription is usually accompanied by a local DNA bending 
\cite{reiss,heu,burl}; the electronic properties  of carbon 
nanotubes drastically depend on their chirality \cite{hamada}; 
T-shape junctions were recently proposed \cite{menon} as nanoscale 
metal-semiconductor-metal contact devices. The geometry manifests itself 
in particular in creation of linear quasi-bound states 
(see e.g. \cite{kircz,gai}). 
On the other hand, the competition between nonlinearity and dispersion 
likewise results into localization of energy. Taken together these 
two localization mechanisms are competitive and one might expect that 
the {\em interplay of geometry and nonlinearity can lead to 
qualitatively new effects}. 
An example of {\em nonlinearity-induced change of the geometry of the 
system} is found in Ref. \cite{dan} where the classical Heisenberg 
model on a two-dimensional manifold was considered. It was shown 
that a periodic topological spin soliton induces a periodic pinch of 
the cylindric manifold. 

\begin{figure}
\centerline{\hbox{
\psfig{figure=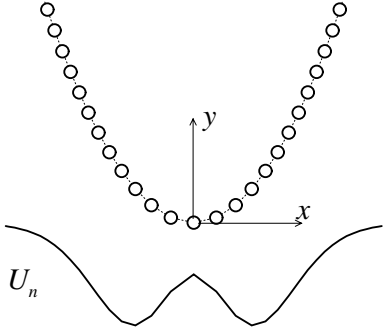,width=50mm,angle=0}}}
\vspace{2mm}
\caption{Curved chain with sites equidistantly placed on a parabola with
the curvature $\kappa=0.5$ at $x=0$ and 
corresponding effective double-well potential $U_n=-\sum_m\,J_{n,m}$}
\label{fig:chain}
\end{figure}

In this paper we consider nonlinear Schr{\"o}dinger (NLS) models 
on a curved one-dimensional manifold and show that the competition 
between curvature and nonlinearity leads to a qualitatively new effect: 
{\em curvature-induced symmetry breaking for non-topological solitary 
excitations}. 
This effect is inherent for both continuum and discrete NLS models. 
We start with the discrete NLS equation 
\begin{eqnarray}
\label{nls}
i \frac{d}{dt} \psi_n +\sum_{m} J_{n m} \psi_m + |\psi_n|^2 \psi_n=0 
\end{eqnarray}
which describes a system of coupled nonlinear oscillators, 
$\psi_n(t)$ being the complex amplitude at the site $n$ ($n=0,\pm 1,\pm 2$). 
In the theory of charge (energy) transfer $\psi_n(t)$ is the wave 
function of the carriers. The Hamiltonian 
\begin{eqnarray}
\label{ham}
H=-\sum_{n} \bigl\{ \sum_{m} J_{n m} \psi^*_n\psi_{m} + 
\frac{1}{2} |\psi_n|^4 \bigr\} 
\end{eqnarray}
and the number of excitations (quanta) $N=\sum_n|\psi_n|^2$ are conserved 
quantities. In Eq. (\ref{nls}) 
the nonlinear term $|\psi_n|^2 \psi_n$ represents a 
self-interaction of the quasiparticle (anharmonicity of vibrations). 
The excitation transfer $J_{n m}$ in the coupling term 
depends on the distance in the embedding space between the sites $n$ and $m$: 
$J_{n m} \equiv J(|\vec{r}_n-\vec{r}_{m}|)$, 
where the radius-vector $\vec{r}_n=(x_n, y_n, z_n)$ characterizes 
the position of the site $n$ on the manifold. The sites are assumed to be
equidistantly placed on plane envelope curve (Fig. \ref{fig:chain}). 
This model is closely related to the so-called wormlike inextensible chain 
model \cite{kr,saito} which is abundantly used in polymer dynamics 
(see {\em e.g.} \cite{kroy,liv,fuk}). 
We consider the properties of nonlinear excitations 
in the vicinity of a smooth bend where the curve 
can be modeled by the parabola without loss of generality 
(as done in Fig. \ref{fig:chain}): 
$y_n=\kappa\,x^2_n/2$ and $z_n=0$ with $\kappa$ being the inverse 
radius of curvature at the bending point. 
The distance between neighboring sites is assumed to be independent on 
the curvature $\kappa$: $|\vec{r}_{n+1}-\vec{r}_n|=1$ 
(inextensibility of the chain). Therefore, 
the excitation dynamics would not depend on the curvature 
of the system in the {\em nearest-neighbor approximation}, when 
$J_{n m}=J\delta_{n-m, \pm 1}$. In many physical problems, however, 
the nearest-neighbor approximation is too crude. For example, the 
DNA molecule contains charged groups, and therefore the 
vibration-excitation transfer is due to the dipole-dipole interaction 
decaying with the distance $r$ as $1/r^3$. Effective dispersive interaction 
between nonlinear layers in nonlinear dielectric superlattices is 
exponential, $\exp(-\alpha r)$, with the inverse radius $\alpha$ 
depending on parameters of the lattice (superlattice spacing, linear 
refractive index and so on) \cite{sup}. 
With this in mind one can see that in general case the number of 
neighbors available for excitation is larger near the bending point 
than far away from it. This gives birth, in particular, to an 
effective double-well potential $U$ (see Fig. \ref{fig:chain} 
and Eq. (\ref{hamc})) which is basically responsible 
for the effects discussed below. 

We investigated both 
%\begin{eqnarray}
%\label{j}
$J_{n m}= J \exp\left(-\alpha |\vec{r}_n- 
\vec{r}_{m}| \right)$
%\end{eqnarray}
and $J_{n m}= J |\vec{r}_n-\vec{r}_{m}|^{-s}$. 
But in this paper we discuss in detail the first case only. 
It is known \cite{magn} that in the case of straight line ($\kappa=0$) 
such NLS model exhibits (for $\alpha < 1.7$) 
bistability in the spectrum of nonlinear stationary states. 
To distinguish these effects from the finite curvature 
effects we use $\alpha=2$.
We study stationary states of the system 
$\psi_n(t)=\phi_n(\Lambda) \, e^{i\Lambda t}$,
where $\Lambda$ is the nonlinear frequency and $\phi_n(\Lambda)$ is 
the amplitude in the $n$-th site. 

%%%%%%%%%%%%%%%%%%%%%%%%%%%%%%%%%%%%%%%%%%%%%%%%%%%%%%%%%%%%%%%%%%%
%\vspace{2mm}
%%%%%%%%%%%%%%%%%%%%%%%%%%%%%%%%%%%%%%%%%%%%%%%%%%%%%%%%%%%%%%%%%%%
%%%%%%%%%%%%%%%%%%%%%%%%%%%%%%%%%%%%%%%%%%%%%%%%%%%%%%%%%%%%%%%%%%%
\vspace{-5mm}
\begin{figure}
\centerline{\hbox{
\psfig{figure=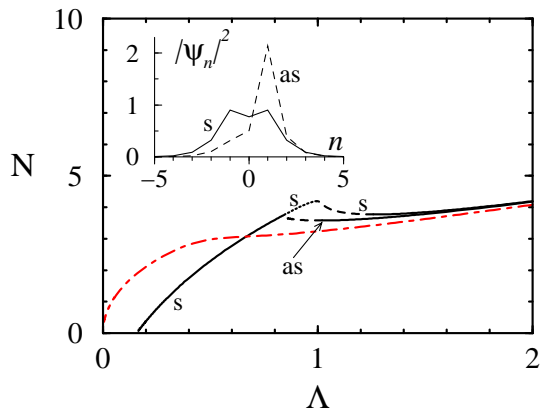,width=70mm,angle=0}}}
%\vspace{0mm}
\caption{The number of excitations $N(\Lambda)$ for the straight 
chain ($\kappa=0$ ; dot-dashed line) and curved chain 
($\kappa=3$ ; solid, dotted, and dashed lines) 
for $\alpha=2$ and $J=6.4$. 
In the inset shapes of the symmetric (s) and asymmetric (as) 
stationary states for $N=3.6$ are shown.}
\label{fig:norm}
\end{figure}
%%%%%%%%%%%%%%%%%%%%%%%%%%%%%%%%%%%%%%%%%%%%%%%%%%%%%%%%%%%%%%%%%%%

Figure \ref{fig:norm} shows $N$ versus $\Lambda$ obtained numerically 
for the straight chain and curved chain. Several new features arise as a 
consequence of the finite curvature:

\begin{itemize}
\item
%%$\bullet$ 
{\em There is a gap} at small $\Lambda$ in which no localized 
excitation can exist. This gap originates from the existence of a 
curvature created {\em linear two-hump localized mode}. 
The gap increases when the curvature $\kappa$ increases. 

\item 
%%$\bullet$ 
There are {\em two branches of stationary states}: a branch of symmetric 
localized excitations (s), which exists for all values of $N$, and a branch 
of {\em asymmetric localized excitations} (as), which 
exists only for $N > N_{th}(\kappa)$. 
The threshold value $N_{th}(\kappa)$ decreases when $\kappa$ 
decreases and vanishes in the limit $\kappa \to 0$.
The symmetric stationary state has a two-humped shape 
(evolved from the linear two-hump localized mode) 
while in the asymmetric state the maximum is shifted either 
to the left or to the right from the center of symmetry $n=0$ 
(see inset in Fig. \ref{fig:norm}). 

\item 
%%$\bullet$ 
For the symmetric stationary states $N(\Lambda)$ is 
{\em non-monotonic} as in Refs. \cite{lr,magn} for NLS models 
on the straight chain with the radius of the dispersive 
interaction above a critical value. Similarly, 
in the case of curved chain there is an interval of $N$ in which three 
different symmetric states coexist for each excitation number. 
Taking into account that for the inverse radius $\alpha=2$ under 
investigation there is no bistability in the straight chain 
(see dot-dashed line in Fig. \ref{fig:norm}), one may conclude 
that {\em bistability is facilitated in the systems 
with finite curvature}.
\end{itemize}

%%%%%%%%%%%%%%%%%%%%%%%%%%%%%%%%%%%%%%%%%%%%%%%%%%%%%%%%%%%%%%%%%%%
\vspace{-3mm}
\begin{figure}
\centerline{\hbox{
\psfig{figure=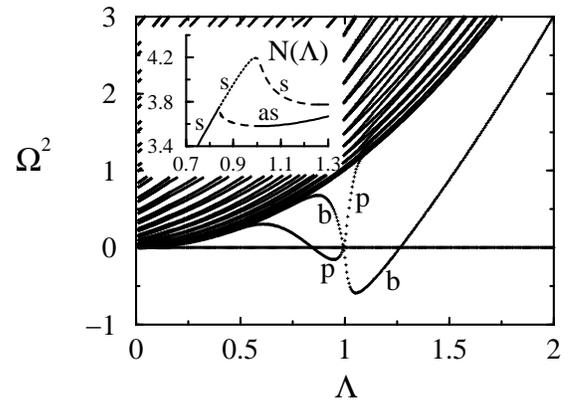,width=70mm,angle=0}}}
%\vspace{0mm}
\caption{Spectrum of linear excitations around the symmetric 
stationary state $\phi_n(\Lambda)$ for $\alpha=2$, $J=6.4$ and 
$\kappa=3$. There are two internal localized modes: Peierls 
antisymmetric (p) and breathing symmetric (b). In the inset the 
dependence $N(\Lambda)$ is shown.}
\label{fig:omega}
\end{figure}
%%%%%%%%%%%%%%%%%%%%%%%%%%%%%%%%%%%%%%%%%%%%%%%%%%%%%%%%%%%%%%%%%%%

The linear stability of the stationary states is investigated 
(as it is described in Refs. \cite{eilb,magn}) letting  
$\psi_n(t)=[\phi_n(\Lambda) +\epsilon_n 
\exp(i\Omega t)] \, \exp(i\Lambda t)$
and linearizing Eq. (\ref{nls}) with respect to $\epsilon_n$. 
By definition the stationary state $\phi_n(\Lambda)$ 
is linearly stable if $\Omega^2>0$. Figure \ref{fig:omega} shows the 
spectrum of linear excitations around the symmetric localized state. 
We find that there are two localized internal modes: Peierls antisymmetric 
mode (p) and breathing symmetric mode (b). There are two intervals of 
instability of the symmetric stationary state: 
when $0.99 < \Lambda< 1.26$ the instability is due to the breathing 
internal mode. In this part of the curve $N(\Lambda)$ (dashed line in 
the inset) $dN/d\Lambda<0$ and the instability agrees with the 
Vakhitov-Kolokolov stability criterion \cite{vk}. 
When $0.84 < \Lambda< 0.99$ the stationary state is unstable with respect 
to the Peierls mode (the corresponding part of $N(\Lambda)$ 
is marked by the dotted line).
In this latter case {\em we observe a violation of 
the Vakhitov-Kolokolov stability criterion} because this 
instability interval corresponds to a part of the curve $N(\Lambda)$ where 
$d N/d \Lambda > 0$ (see Fig. \ref{fig:norm}).

%%%%%%%%%%%%%%%%%%%%%%%%%%%%%%%%%%%%%%%%%%%%%%%%%%%%%%%%%%%%%%%%%%%
%\vspace{2mm}
\begin{figure}
\centerline{\hbox{
\psfig{figure=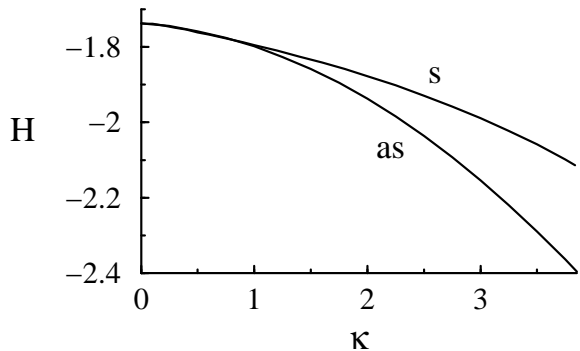,width=75mm,angle=0}}}
%\vspace{2mm}
\caption{Energy of symmetric (s) and asymmetric (as) nonlinear 
excitations versus the curvature $\kappa$ for $\alpha=2$, 
$J=6.4$, and $N=4$.}
\label{fig:energy}
\end{figure}
%%%%%%%%%%%%%%%%%%%%%%%%%%%%%%%%%%%%%%%%%%%%%%%%%%%%%%%%%%%%%%%%%%%
%%%%%%%%%%%%%%%%%%%%%%%%%%%%%%%%%%%%%%%%%%%%%%%%%%%%%%%%%%%%%%%%%%%
\vspace{-15mm}
\begin{figure}
%\centerline{\hbox{
\hspace{-33mm}
\psfig{figure=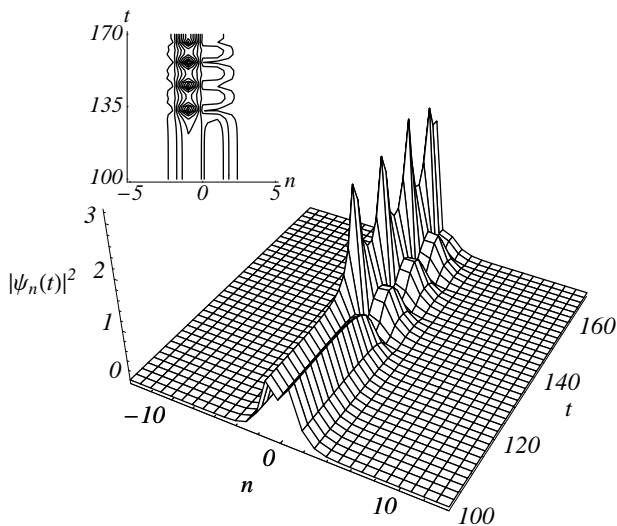,width=100mm,angle=270}
%}}
\vspace{-40mm}
\caption{Switching from the symmetric state to the asymmetric state at 
$\alpha=2$, $J=6.4$, $\kappa=4$, and $N=3.96$.}
\label{fig:dynamics}
\end{figure}
%%%%%%%%%%%%%%%%%%%%%%%%%%%%%%%%%%%%%%%%%%%%%%%%%%%%%%%%%%%%%%%%%%%

The energy of the symmetric as well as asymmetric 
localized states is monotonically decreasing function 
of the curvature (see Fig. \ref{fig:energy}). Thus, one may 
expect that the {\em nonlinear localized excitation may facilitate 
bending of flexible molecular chain}. 
For $N > N_{th}(\kappa)$, when symmetric and asymmetric 
stationary states coexist, the {\em asymmetric state is always 
energetically more favorable}. 
Taking into account that in the linear limit the ground state of the model 
is symmetric with respect to the center of symmetry $n=0$ one can conclude 
that {\em combined action of the finite curvature and nonlinearity provides 
the symmetry breaking} in the system.
A typical evolution of the curved NLS model in the case when the symmetric 
stationary state becomes unstable is shown in Fig. \ref{fig:dynamics}. 
One can see that the two-humped initial symmetric state evolves 
into an asymmetric state with an excited breathing internal mode. 

Apart from the smooth bending we studied the case of the edge: 
$y_n=|x_n| \,\tan(\theta) \, $. Qualitatively the same results 
were obtained for this case, too.

%%%%%%%%%%%%%%%%%%%%%%%%%%%%%%%%%%%%%%%%%%%%%%%%%%%%%%%%%%%%%%%%%%%
%\vspace{0mm}
\begin{figure}
\centerline{\hbox{
\psfig{figure=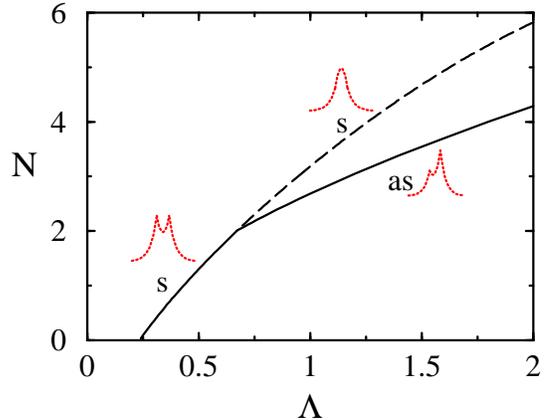,width=70mm,angle=0}}}
\vspace{-1mm}
\caption{The dependence $N(\Lambda)$ for symmetric (s) and asymmetric 
(as) nonlinear excitations in the two-impurity minimal model 
(\protect\ref{toy}) with $\epsilon=0.7$ and $a=1$.}
\label{fig:toy}
\end{figure}
%%%%%%%%%%%%%%%%%%%%%%%%%%%%%%%%%%%%%%%%%%%%%%%%%%%%%%%%%%%%%%%%%%%

To clarify the physical meaning of obtained results we 
pass to the continuum limit of Eq. (\ref{ham}) where the Hamiltonian 
$H$ can be represented as
\begin{eqnarray}
\label{hamc}
H=\int_{-\infty}^{\infty} \Biggl( \frac{1}{2} L^2(\ell)
\Bigl| \frac{\partial \psi}{\partial \ell} \Bigr|^2+
U(\ell)|\psi|^2 - \frac{1}{2} |\psi|^4 \Biggr) d \ell \; . 
\end{eqnarray}
Here $\ell$ is the arclength (the continuum analogue of $n$), 
$L^2(\ell)= \int_{-\infty}^{\infty}(\ell'-\ell)^2\,J_{\ell,\ell'} 
d \ell'$ 
is the square of the  dispersive length of the excitation, and 
$U(\ell) =  -\int_{-\infty}^{\infty}\,J_{\ell,\ell'} d \ell'$ is the 
energy shift due to the coherent interaction.
The inverse length $L^{-1}(\ell)$ and attractive potential $U(\ell)$ 
are generally 
double-well functions. At the bottom of Fig. \ref{fig:chain} 
the discrete analogue of $U(\ell)$, 
$U_n=-\sum_m\,J_{n, m}$, is shown.
In the vicinity of the bend there are two 
symmetric sections of the chain in which the number of neighbors entering 
into the range of the dispersive interaction, $~1/\alpha~$, is higher than 
anywhere else on the chain. In this way a double-well potential structure 
results. The positions and depths of the 
wells are determined by the range of the dispersive interaction 
and the peculiarities of the bending geometry. 
E.g., in the case of the edge, $y_n=|x_n| \,\tan (\theta) \,$, we 
get for small $\theta$  \\ $L^2(\ell)\approx 
 \frac{J}{\alpha^3}\left(4+((\alpha^2 |\ell|+1)^2+1)
(e^{-\alpha|\ell|\cos(2\theta)}-e^{-\alpha|\ell|})\right)$ \\
and $U(\ell)\approx -\frac{J}{\alpha}\left(2+
e^{-\alpha|\ell|\cos(2\theta)}-e^{-\alpha|\ell|}\right)$. \\
It is worth noting that the 
Hamiltonian similar to Eq. (\ref{hamc}) with $U(\ell) \equiv 0$ was 
introduced in Ref. \cite{peyrard} to study the effects of the 
DNA bending on breather trapping. The step function dependence 
of the dispersive length $L(\ell)$ was there postulated. In all cases 
bending manifests itself as a trap for nonlinear excitations. 
To obtain an analytical solution of the problem 
one can approximate $L(\ell)$ by a constant and $U(\ell)$ by two 
$\delta$-functions. Then we arrive at the following minimal model
\begin{eqnarray}
\label{toy}
i\,\frac{\partial \psi}{\partial t} + 
\frac{\partial^2 \psi}{\partial \ell^2} + 
\epsilon (\delta(\ell-a)+\delta(\ell+a)) \psi + 
|\psi|^2 \psi=0 \; , 
\end{eqnarray}
where the parameter $\epsilon$ is the depth of the well. Introducing 
into Eq. (\ref{toy}) the stationary state ansatz
$\psi(\ell,t)=\phi(\ell) \exp(i\Lambda t)$ we find that 
the corresponding nonlinear eigenvalue problem can be easily reduced 
to an algebraic one. 
The results of the analytical consideration are shown in Fig. 
\ref{fig:toy}. The stationary state (s) is unique and symmetric for small 
number of quanta $N$. But when the number of quanta exceeds some 
critical value there are two stationary states: symmetric (s) and asymmetric 
(as), with the asymmetric state being energetically favorable. Thus, 
similarly to the original model, the minimal model demonstrates a symmetry 
breaking effect (localization in one of the wells). 

We conclude that in the NLS model the interplay of 
curvature and  nonlinearity leads to the symmetry breaking when the 
asymmetric stationary state becomes energetically favorable than 
the symmetric one. We have found that the energy of localized states decreases 
with the increasing of curvature. One may expect that the nonlinear 
localized excitation may facilitate bending of a flexible molecular 
chain. In all cases bending manifests itself as a trap for nonlinear 
excitations. For the curved NLS model in the case of the 
physically important dipole-dipole dispersive interaction 
$J_{n m} = |\vec{r}_n-\vec{r}_{m}|^{-3}$ qualitatively the 
same properties are found. Using as an example the 
Davydov-Scott model \cite{dav,scott} with the dipole-dipole coupling 
between sites \cite{leonor} 
for the parameters which characterize the motion of the amide-I excitation 
in proteins (they correspond to $N=0.64$ in our model) decreasing of the 
ground state energy due to a trapping by the bending with $\kappa=1$ 
($\kappa=3$) comprise $10\%$ ($46\%$) of the whole excitation energy. 
This may be also important for the dynamics of DNA molecule 
where as was mentioned above the vibration-excitations are described by 
Eq. (\ref{nls}) with the dipole-dipole dispersive interaction \cite{ming}. 
Finally, a violation of the Vakhitov-Kolokolov stability 
criterion in the case when the instability is due to the softening 
of the Peierls internal mode is found.

%\bigskip
%%%%%%%%%%%%%%%%%%%%%%%%%%%%%%%%%%%%%%%%%%%%%%%%%%
\section*{Acknowledgments}
%%%%%%%%%%%%%%%%%%%%%%%%%%%%%%%%%%%%%%%%%%%%%%%%%%
We thank A.C.~Scott for the helpful discussions. 
Yu.B.G. and S.F.M. would like to express their thanks 
to the Department of Mathematical Modelling,
Technical University of Denmark where the major 
part of this work was done. 

%%%%%%%%%%%%%%%%%%%%%%%%%%%%%%%%%%%%%%%%%%%%%%%%%%

%%%%%%%%%%%%%%%%%%%%%%%%%%%%%%%%%%%%%%%%%%%%%%%%%%
\end{multicols}
\end{document}